\input{aipcheck}

\documentclass[
    ,final 
  ]
  {aipproc}

\layoutstyle{8x11double}

\begin{document}

\title{Gauges, propagators, and physics}

\classification{11.15.Ha; 11.15.Tk; 12.38.Aw; 12.38.Gc; 12.38.Mh}
\keywords      {Gauge fixing; Correlation functions; Functional methods; Lattice}

\author{Axel Maas}{
  address={Karl-Franzens-University Graz, Universit\"atsplatz 5, A-8010 Graz, Austria}
}

\begin{abstract}
When a theory shall be described at all scales, it is necessary to start from its elementary degrees of freedom. Herein, one possible chain of steps for this purpose will be briefly outlined for the example of a gauge theory, like QCD. Starting with the elementary constituents, gluons and quarks, step by step  the final observables, physical states, will be built up. This process is based on the elementary correlation functions, and uses a combination of both numerical lattice calculations and functional continuum methods. While the process uses a fixed gauge at each intermediate step, the final observables are gauge-invariant.
\end{abstract}

\maketitle

\section{Why a gauge-fixed approach?}

In principle, it is possible to calculate any observable quantity in a gauge theory without fixing a gauge. However, for essentially all practical calculations in non-integrable gauge theories it is useful to fix a gauge at least for some of the intermediate steps for at least two reasons. On the one hand, many calculations simplify considerably when choosing a suitable gauge. Especially perturbation theory provides a rich set of examples for this fact. On the other hand, fixing the gauge permits to construct intermediate results, like the correlation functions of the elementary degrees of freedom. These can then be used to construct observables out of them. This is, e.\ g., done in Bethe-Salpeter or Faddeev equations for hadronic properties \cite{Alkofer:2000wg}. Furthermore, they can be used to exchange results between different methods, e.\ g., lattice and functional methods \cite{Fischer:2008uz}. The latter case is particularly important if a combination of methods is applied to solve a certain problem which cannot be solved using a single method. A pertinent example is to use functional methods to extend lattice results from the narrow energy domain accessible in numerical simulations to the energy reach of at least 9 orders of magnitude typical for the standard model.

\section{Gauges}

In a gauge theory, the elementary degrees of freedoms cannot be described without fixing a gauge. E.\ g., in QCD these are the quarks and gluons. Fixing the gauge is merely a choice of coordinate system in the field configuration space. This choice is often facilitated by the introduction of additional auxiliary fields, the so-called ghost fields. However, gauge theories based on non-Abelian gauge groups introduce additional complications in the process of gauge-fixing, due to the so-called Gribov-Singer ambiguity. A short description of the inherent problems can be found in \cite{Alkofer:2000wg,Maas:2010wb}. An important consequence of this ambiguity is that in general a perturbatively unique gauge description becomes a family of gauges non-perturbatively.

One such family which has been rather intensively studied is the Landau gauge family, and herein in particular the subset of so-called first-Gribov-region Landau gauges \cite{Maas:2010wb}. The reason for the choice of Landau gauge is its technical advantages, for details see \cite{Fischer:2008uz}. However, many other gauges have been studied as well, see e.\ g.\ \cite{Alkofer:2000wg} for an introduction. It should be noted that it is rather non-trivial to choose the same gauge with different methods beyond perturbation theory, see \cite{Fischer:2008uz,Maas:2010wb,Maas:2009se}. In the remainder, it will be assumed that the construction proposed in \cite{Maas:2010wb,Maas:2009se} is possible.

Once a choice of gauge is done, the elementary degrees of freedom are well defined. It is then possible to determine their correlation functions, in particular their propagators and vertices. This will be done here using both lattice and functional continuum methods.

\section{Propagators}

\begin{figure}[!ht]
 \begin{tabular}{c}
  \includegraphics[width=0.4\textwidth]{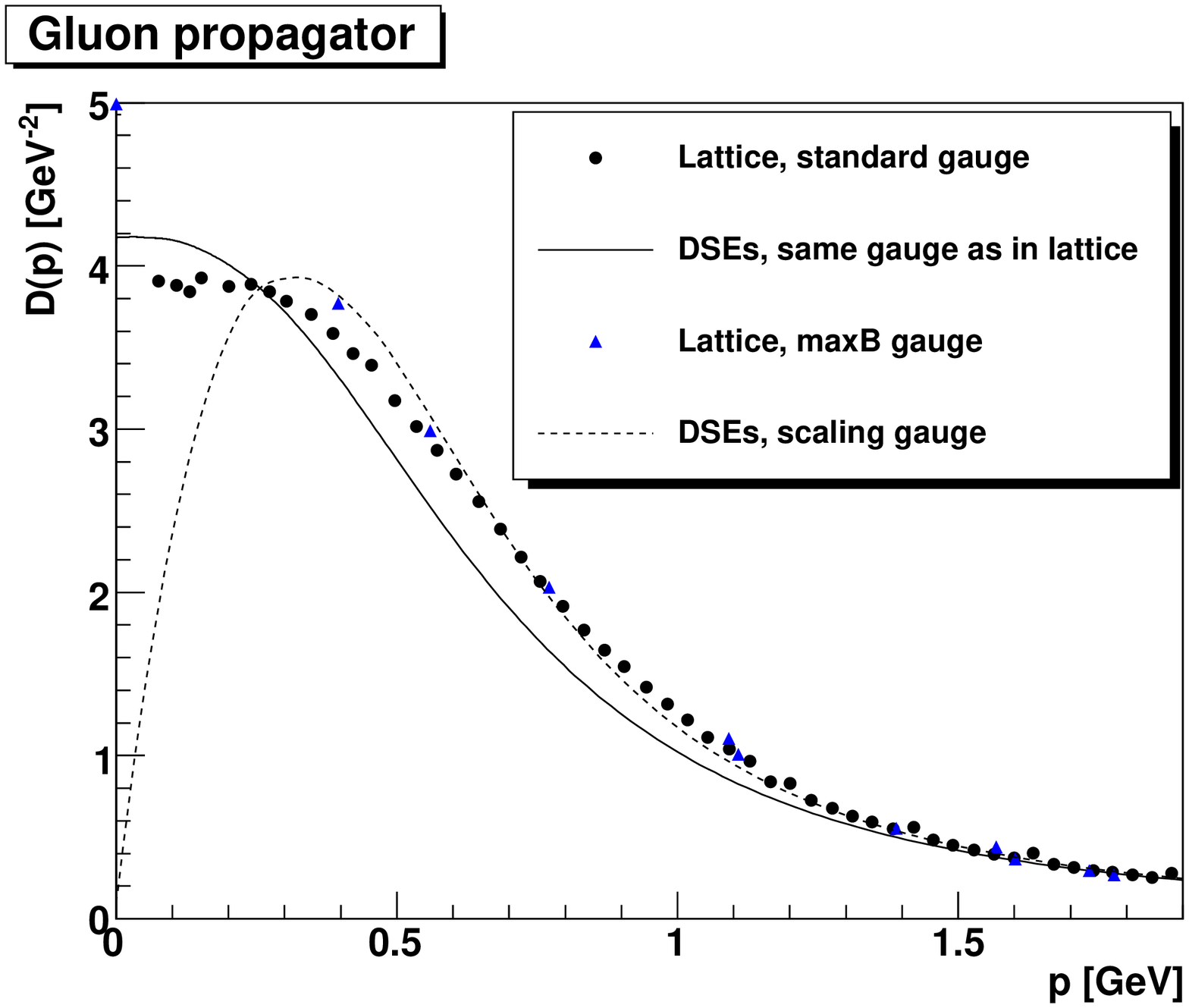}\cr

  \includegraphics[width=0.4\textwidth]{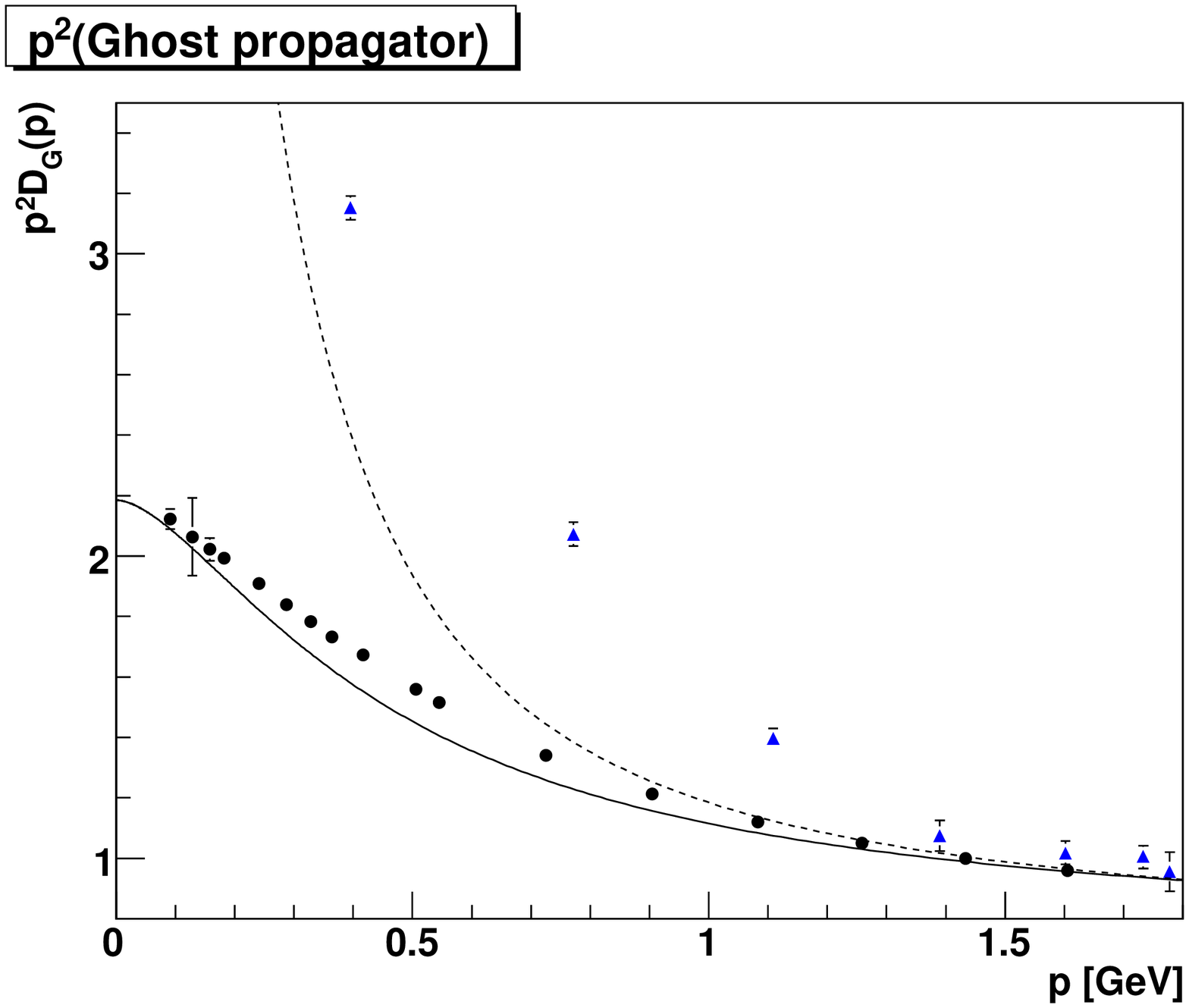}\cr

  \includegraphics[width=0.4\textwidth]{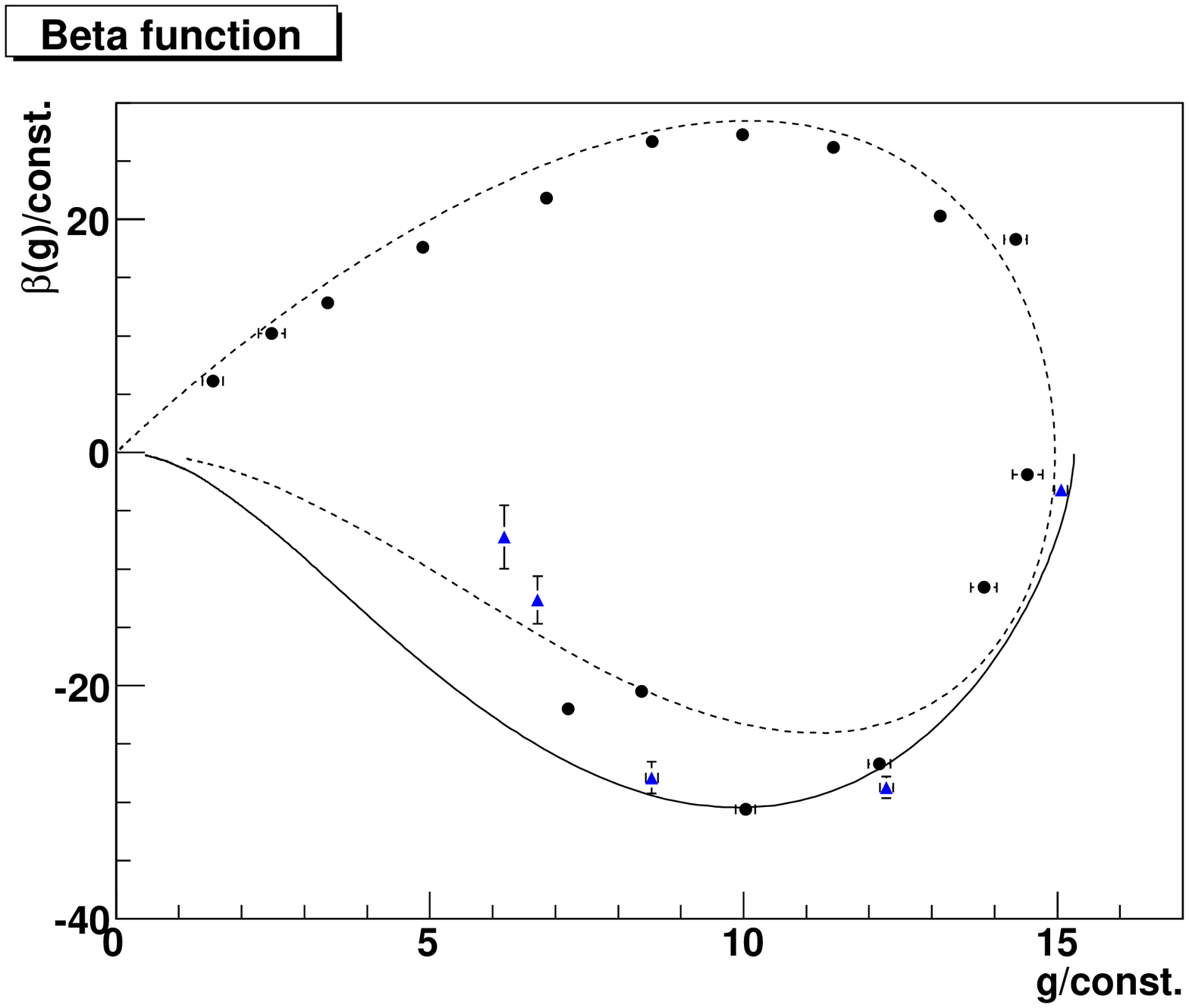}\cr
 \end{tabular}
  \caption{\label{figym}The gluon propagator (top panel), the ghost dressing function (middle panel) and the $\beta$-function (bottom panel, axes rescaled) for lattice calculations using a standard gauge \protect\cite{Bogolubsky:2009dc}, its reproduction using Dyson-Schwinger equations (DSEs) \protect\cite{Fischer:2008uz}, as well as DSE results for a scaling gauge \protect\cite{Fischer:2008uz} and for a third gauge in lattice calculations, the maxB gauge \protect\cite{Maas:2009se,Maas:unpublished}. Concerning the assignment of the gauges, see \protect\cite{Maas:2010wb}.}
\end{figure}

\begin{figure}
  \includegraphics[width=0.4\textwidth]{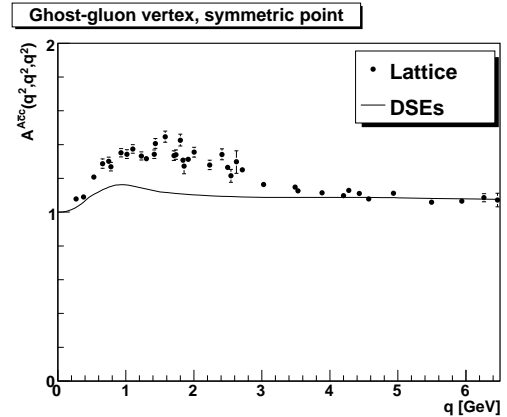}
  \caption{\label{figv} The single tensor structure of the ghost-gluon vertex from lattice calculations in the standard gauge \cite{Cucchieri:2008qm} and from DSEs in the scaling gauge \cite{Schleifenbaum:2004id} at the symmetric point, i.\ e., for all momenta of the same magnitude \cite{Schleifenbaum:2004id,Cucchieri:2006tf}.}
\end{figure}

Once the gauge is fixed, it is possible to determine both propagators \cite{Fischer:2008uz,Cucchieri:2006tf} as well as higher vertex functions \cite{Cucchieri:2006tf,Alkofer:2008tt}, though this can be rather challenging in detail. Results for the propagators of the gluon, the ghost, and the $\beta$-function of Yang-Mills theory for various gauges and methods are shown in figure \ref{figym}. It should be noted that the renormalization scheme has been chosen differently for the different gauges for convenience, and therefore the $\beta$-functions do not agree for the different gauges. That this is also possible at the level of vertices is shown in figure \ref{figv} for the ghost-gluon vertex.

The results show that it is possible to obtain rather well agreeing results with functional and continuum calculations, at least on the level of propagators \cite{Fischer:2008uz}. Thus, it is possible to determine the correlation functions, which can then be used to obtain physics results.

\section{Physics}

The main aim of this combination of methods is to obtain results which cannot be easily determined with either method alone. Particular examples are chiral limit physics \cite{Fischer:2006ub} or the finite density region in the QCD phase diagram \cite{Nickel:2006vf}. Useful quantities in this context are meson masses \cite{Alkofer:2000wg,Fischer:2006ub} and order parameters, like the Polyakov loop \cite{Fischer:2010fx,Braun:2007bx,Braun:2009gm}.

\begin{figure}
 \begin{tabular}{c}
  \includegraphics[width=0.4\textwidth]{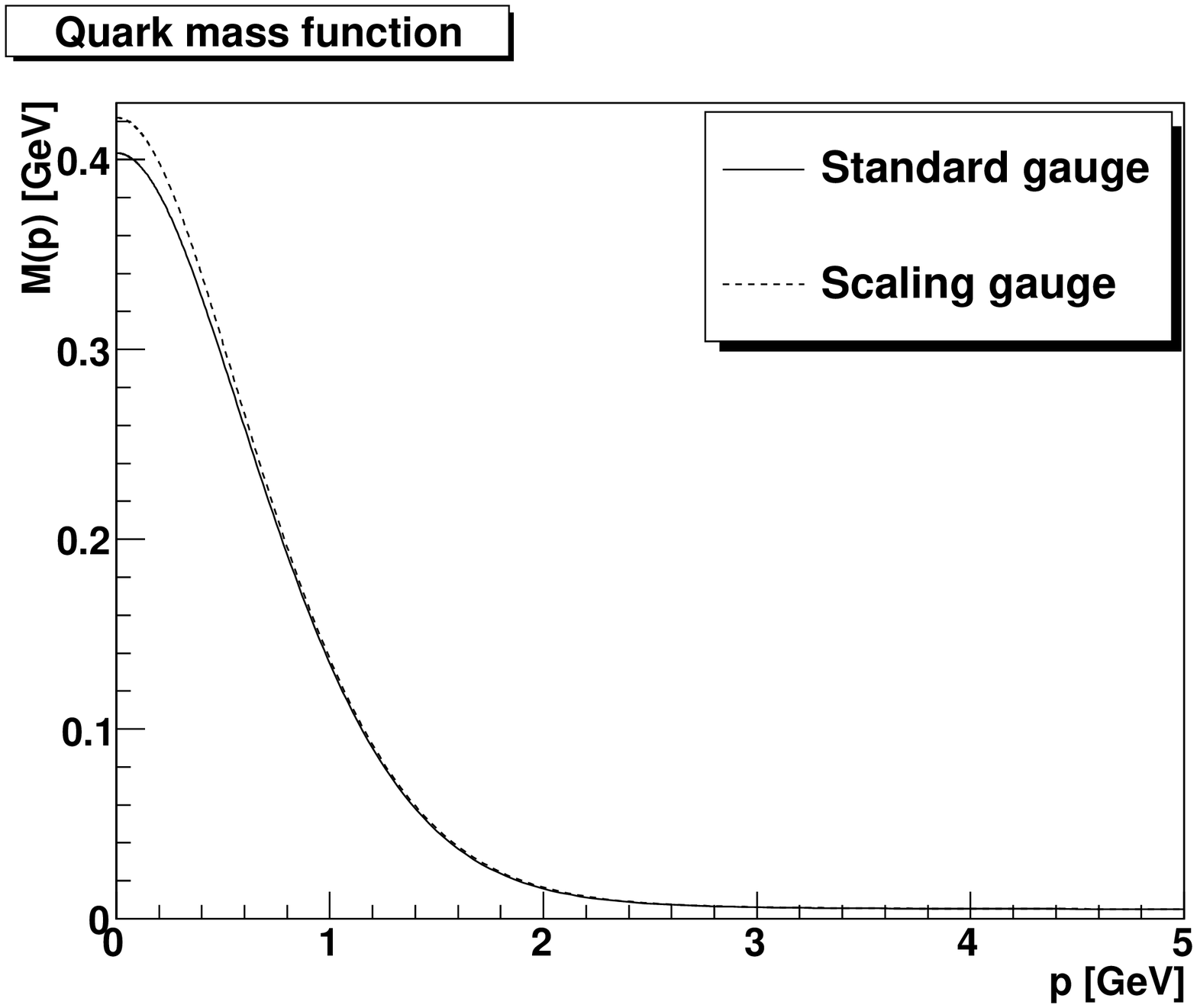}\cr
  \includegraphics[width=0.4\textwidth]{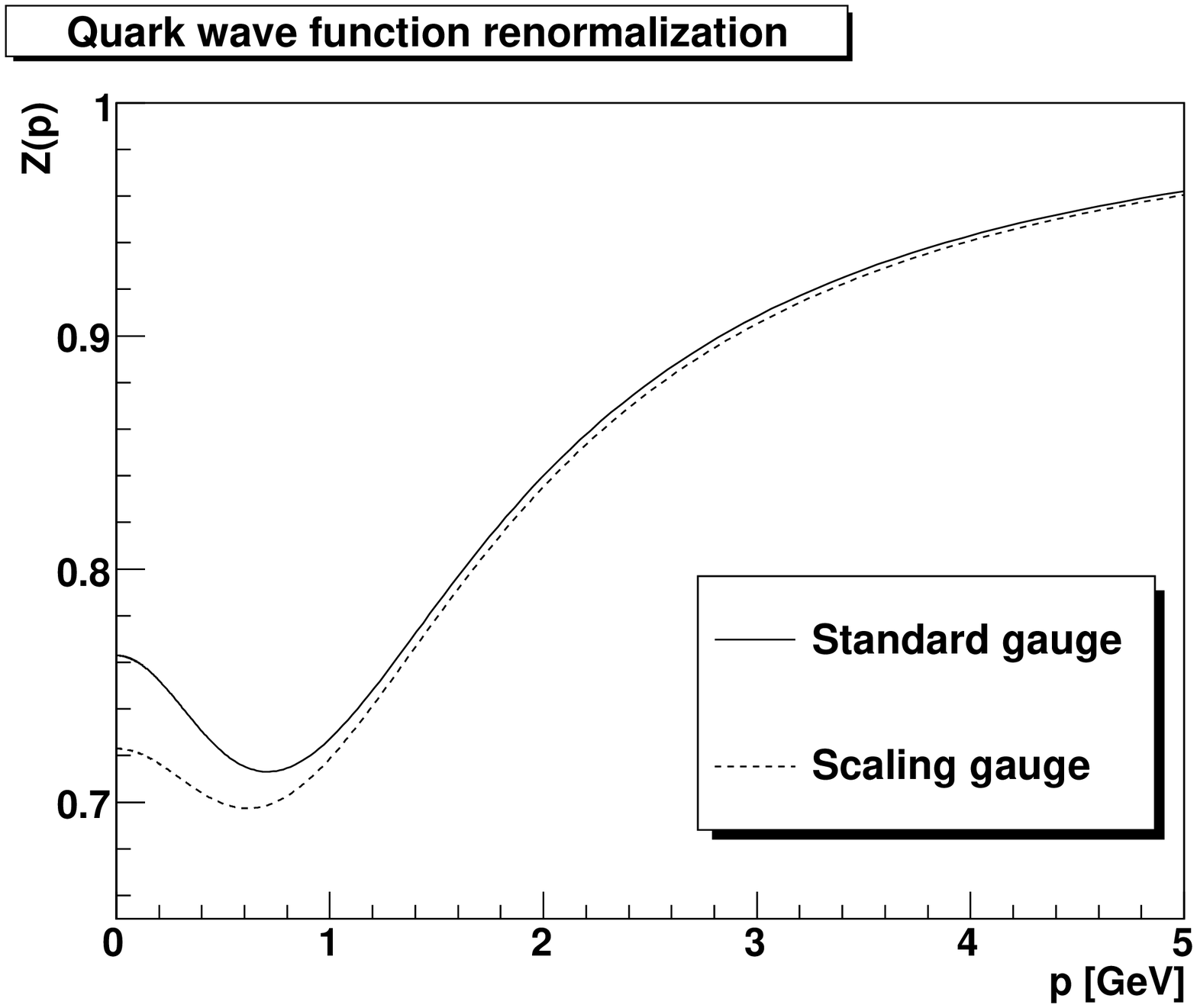}\cr
\end{tabular}
  \caption{\label{figqm} The quenched quark mass function (top panel) and wave function renormalization (bottom panel) for the standard gauge choice and the scaling gauge used in figure \ref{figym} from Dyson-Schwinger equations \cite{Blank:2010pa}.}
\end{figure}

As an example, consider the mass of a meson \cite{Fischer:2006ub,Blank:2010pa,Fischer:2005en}. For simplicity take the quenched case. Then, the mass of the meson can be determined using the Bethe-Salpeter equation \cite{Alkofer:2000wg}, which requires further input in form of the quark propagator and the quark-gluon vertex. This propagator can be determined again using the same methods, and will depend on the chosen gauge \cite{Fischer:2006ub,Blank:2010pa,Aguilar:2010cn}. Results for different gauges for its two independent dressing functions are shown in figure \ref{figqm}.

The second ingredient necessary is the quark-gluon vertex. It can be either calculated self-consistently \cite{Alkofer:2008tt}, or it can be parametrized using the gauge-invariance of meson masses to fix the parameters \cite{Blank:2010pa}. Fixing the vertex using the pion properties with the latter method, it is possible to calculate, e.\ g., the $\rho$-meson properties \cite{Blank:2010pa}. Since the gauge dependence is, up to truncation artifacts, canceled between the different contributions from the gluon, the quark, and the quark-gluon vertex, the $\rho$-meson properties then also agree with the physical values for physical quark masses \cite{Blank:2010pa}. On the other hand, if this were not the case, this would hint towards severe truncation artifacts, and such problems could then be used to improve the truncations. This approach can be supplemented by more general considerations, as in QED \cite{Kizilersu:2009kg}. However, the particulars of the non-Abelian case unfortunately makes this somewhat more intricate \cite{vonSmekal:1997vx}.

\section{Concluding remarks}

By using gauge-fixed correlation functions as a common basis it is possible to combine different genuine non-perturbative methods in such a way as to extend their reach to problems none of them could reliably access alone. The basis for this combination has been constructed during the recent years. With these technologies in place, this approach can now be applied to increasingly complicated problems. Prominent examples are light, dynamical fermions \cite{Fischer:2006ub} and the QCD phase diagram at both finite temperature and density \cite{Nickel:2006vf,Fischer:2010fx,Braun:2007bx,Braun:2009gm}. Furthermore, first investigations have started to also extend their reach to physics question beyond QCD, in particular the electroweak sector in and beyond the standard model \cite{Maas:unpublished,Aguilar:2010cn,Maas:2010nc,Sannino:2009za}. Thus, this approach to treat gauge theories beyond perturbation theory is one of the possible avenues to study strongly interacting systems at all scales.

\begin{theacknowledgments}
 This work has been supported by the FWF under grant number M1099-N16.
\end{theacknowledgments}

\bibliographystyle{aipproc}
\bibliography{bib}

\end{document}